\documentclass[prl,twocolumn,showpacs]{revtex4}
\usepackage{amssymb}
\usepackage{amsmath}
\usepackage{epsfig}
\begin{document}
\title{Nature of intrinsic relation between Bloch-band tunneling and modulational instability}
\author{V.A. Brazhnyi$^1$, V.V. Konotop$^{1}$, and V. Kuzmiak$^2$}
\affiliation{$^1$Centro de F\'{\i}sica Te\'{o}rica e Computacional, 
Universidade de Lisboa,
 Complexo Interdisciplinar, Avenida Professor Gama Pinto 2, Lisboa 
1649-003, Portugal\\
$^2$ Institute of Radio Engineering and Electronics, Czech
Academy of Sciences, Chaberska 57, 182 51 Prague 8, Czech Republic}
%\date{ }
\begin{abstract}
On examples of Bose-Einstein condensates embedded in two-dimensional optical lattices we show that in 
nonlinear periodic systems modulational instability and inter-band tunneling are intrinsically related 
phenomena. By direct numerical simulations we found that tunneling results in attenuation or enhancement of instability. On the other hand, instability results in asymmetric nonlinear tunneling. The effect strongly depends on the band gap structure and it is 
especially significant in the case of the resonant tunneling.  The symmetry of the coherent structures 
emerging from the instability reflects the symmetry of both the stable and the unstable states between 
which the tunneling occurs. Our results provide an evidence of profound effect of the band structure 
on superfluid-insulator transition. 
\end{abstract}

\maketitle

Bose-Einstein condensate (BEC) embedded in an optical lattice represents a remarkable laboratory 
for study of the interplay among various physical phenomena, having general physical nature. One 
of such phenomena-- the modulational instability -- is a characteristic feature of nonlinear 
systems~\cite{Abdullaev}. For BECs (where it is also referred to as dynamical instability) it has 
been described theoretically in ~\cite{WuNiu,instability,BKS,MPS} and observed experimentally in 
~\cite{intsab-experim}. Another phenomenon, which is typical for linear periodic systems, is the 
Landau-Zener tunneling. It was also discussed in the context of the BEC applications~\cite{Landau-Zener,KKS} 
and explored in a series of experiments~\cite{Morsch}, where the phenomenon was implemented in 
one-dimensional (1D) optical lattices and was stimulated by the effective linear force created by the 
lattice acceleration. The existing theoretical descriptions, based on two-level models, however, 
are not unanimous about how Landau-Zener tunneling occurs, except one point: 
two-body interactions (i.e. the nonlinearity) make the tunneling asymmetric. It was suggested in Ref.~\cite{KKS}, that asymmetry stems from the fact that the Landau-Zener tunneling in a nonlinear system is intimately related to the instability. Indeed, tunneling couples two band edges with the 
eigenstates that acquire different stability properties in the presence of the nonlinearity, one of 
them being unstable. In this case the initial atomic distribution turns out to be of crucial importance.  
Recently, Landau-Zener tunneling and modulational instability  have been explored within the unique 
experimental setting~\cite{experiment}. However, the data available so far, are not conclusive about 
the reasons for asymmetry of nonlinear Landau-Zener tunneling. Among the other reasons, it is due to 
the fact that to stimulate the tunneling a linear force is necessary. Such a force gives rise to other 
phenomena, such as, for example, Bloch oscillations \cite{Morsch}, which in turn, can also imply the asymmetry.

>From the point of view of understanding the fundamental relation between tunneling and modulational instability, flexibility of multidimensional lattices in general~\cite{Greiner}, and 2D lattices, 
in particular, open new perspectives, offering diversified possibilities of the observing tunneling  and instabilities, without any use of external forces, namely, that of the lattice acceleration. For systems of cold noninteracting atoms this has already been discussed in~\cite{Kolovsky}. Moreover, within the framework of a different system, which is a beam propagation through an optically induced 2D lattice, nonlinear tunneling has been observed experimentally in~\cite{kivshar}. 
In what follows tunneling without external forces is referred to as Bloch-band tunneling and is subject of our principal concern. 

Our main goal is to explore intrinsic relation between Bloch-band tunneling and modulational instability. We show that tunneling can either enhance of attenuate instability, as well as instability can result in asymmetry of tunneling with respect to initial population of stable and unstable states.
Experimentally such states can be created by manipulating with noninteracting atoms and subsequent switching in the nonlinearity, what can be managed by means of Feshbach resonance. We argue that the band 
structure is the most relevant characteristic governing the dynamics and supports resonant 
exchange of the atoms between bands. As an important physical application of our results, we notice that instability of Bloch waves, with subsequent emergence of coherent 
localized structures, implies the superfluid-insulator transition (see e.g.~\cite{WuNiu}, as well as \cite{superfluid} for discussion in the tight-binding approximation, and \cite{fluid-exper} for experimental 
observation). We thus provide an evidence of profound influence of the band structure on the superfluid-insulator transition and suggest a possibility of its managing by simple control of parameters of an optical lattice. 
  
We consider an effectively 2D BEC, with a positive scattering length, governed by the Gross-Pitaevskii 
equation 
\begin{equation}
\label{GPE}
 i\frac{\partial\Psi}{\partial t} =-\Delta\Psi+V({\bf r})\Psi + |\Psi|^2\Psi\,.
\end{equation}
Here time is measured in the units of $2\hbar/E_R$, where $E_R=\hbar^2k^2/(2m)$ is the recoil energy, 
$k=2\pi/\lambda$ is the wave vector of the laser beams considered equal in all directions, 
and coordinates ${\bf r}=(x,y)$ are measured in the units of $k^{-1}$. 
We restrict our consideration to the potential
\begin{eqnarray}
\label{OL}
V({\bf r}) = V_0\left[\cos(2x) + \cos(2y) - \cos(2x)\cos(2y)\right]
\end{eqnarray}
where $V_0$ is measured in the units of $2E_R$.  

\begin{figure}[h]
\epsfig{file=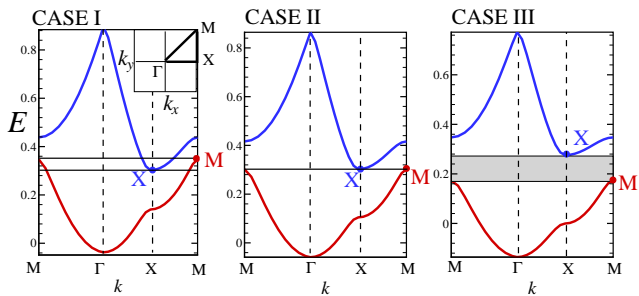,width=8.5cm}%,angle=-90}
\vspace{1. true cm}
\caption{[Color online] The two lowest bands of the potential (\ref{OL}).
In the Case~I, where $V_0=-0.2$, the bands overlap: $E_X-E_M\approx -0.045$. When $V_0=-0.2575$ the band 
structure transforms into that shown in the Case~II: the gap edges  acquire practically the same value: 
$E_X-E_M \approx 10^{-5}$. Further increase of $|V_0|$ leads to opening the gap, as is shown in Case~III, 
where $V_0=-0.27$ and $E_X-E_M \approx 0.1025$. The inset in Case~I shows notations of the high symmetry 
points in the Brillouin zone.}
\label{fig1}
\end{figure}

When the nonlinearity is weak enough and the lattice amplitude is of order of the recoil energy, the 
underlying band structure of $V({\bf r})$ appears to be the most relevant characteristic. Then the 
problem can be described in the effective mass approximation~\cite{instability,BKS}. We start with 
the situation where a band structure possesses a full gap in all directions (Case~III in Fig.~\ref{fig1}). 
The respective linear system does not allow tunneling between bands unless additional external factors 
are involved. The nonlinearity couples modes in different bands making possible local transfer atoms 
among them~\cite{KKS,kivshar}.

When Bloch bands overlap in different directions (Case~I in Fig.~\ref{fig1}), the linear  tunneling between 
states with the same energy is possible.  It turns out, however, that the most intensive exchange of atoms 
between the bands occurs not in this case, but in the case of a {\em resonant tunneling}, when matching 
conditions are satisfied, i.e. when four waves satisfy simultaneously the energy and momentum conservation 
laws (see e.g.~\cite{Maimistov}). Such a lattice can be created by adjusting the laser intensity as it is shown in Case II of Fig.~\ref{fig1}. The states in points M (of the first band) and X (of the second 
band) possess not only the same energy $E_{\rm M}=E_{\rm X}$ [hereafter 
$E_{\alpha}=E({\bf q}_{\alpha})$ with $E({\bf q})$ being the eigenvalue of the operator $-\Delta+V(x,y)$; explicit forms of the dependence 
$E({\bf q})$ are shown in Fig.~\ref{fig1}], but also satisfy the conservation of the 
quasi-momentum. By denoting an arbitrary vector of the reciprocal lattice by ${\bf Q}$, the last condition 
can be expressed either as ${\bf q}_{\rm M}={\bf q}_{\rm X}+{\bf q}_{\rm X}-{\bf q}_{\rm M}+{\bf Q}$ if one 
considers tunneling from initial state X to the state M   or as 
${\bf q}_{\rm X}={\bf q}_{\rm M}+{\bf q}_{\rm M}-{\bf q}_{\rm X}+{\bf Q}$ if one considers tunneling 
from M to X. A process where the majority of particles is initially concentrated in the state A from where they tunnel into the state B will be referred to as A$\to$B tunneling.

Resonant wave interactions available in Cases I and III, are those conventionally referred 
to as self-phase and cross-phase modulations and governed by the integrals $\chi_{_{X,M}}=4\int_{V_0} \Psi_{\rm X,M}^4 d{\bf r}$ 
and $\chi=4\int_{V_0} \Psi_{\rm X}^2\Psi_{\rm M}^2 d{\bf r}$, respectively [hereafter $\Psi_{\rm M,X}$ are the Bloch states, real for the points X and M, and $V_0=(0,\pi)\times(0,\pi)$ is a volume of the unit cell of the lattice]. In the two-mode approximation one can introduce slowly 
varying modulations ${\cal A}_{\rm X,M}$ of the states $\Psi_{\rm M,X}$ and represent (in the leading order) 
\begin{eqnarray}
\label{psi}
\Psi={\cal A}_{\rm X}({\bf r},t)\Psi_{\rm X}({\bf r})e^{-iE_Xt}+
{\cal A}_{\rm M}({\bf r},t)\Psi_{\rm M}({\bf r})e^{-iE_Mt}\,.
\end{eqnarray}
Then, following the standard steps (see e.g.~\cite{BKS}), one deduces the Hamiltonian governing the dymnamics slow amplitudes ($\partial A_\alpha/\partial t=-i\delta H/\delta \bar{A}_\alpha$):
\begin{eqnarray}
\label{ham1}
H=\frac 12 \int d{\bf r} \left\{\left|\hat{M}_{\rm X}^{-1}\nabla {\cal A}_{\rm X}\right|^2+
\left|\hat{M}_{\rm M}^{-1}\nabla {\cal A}_{\rm M}\right|^2
\right.
\nonumber 
\\
\left.
+\chi_{_{\rm X}}\left|{\cal A}_{\rm X}\right|^4+\chi_{_{\rm M}}\left|{\cal A}_{\rm M}\right|^4+
4\chi \left|{\cal A}_{\rm X}\right|^2 \left|{\cal A}_{\rm M}\right|^2\right\}.
\end{eqnarray}
Here $\hat{M}_{\alpha}^{-1}$ ($\alpha=$X,M) is a $2\times 2$ tensor of the inverse effective mass in a point ${\bf q}_\alpha$ whose entries are defined by $M^{-1}_{ij}({\bf q}_\alpha)={\partial^2 E({\bf q}_\alpha)}/{(\partial q_{\alpha,i}\partial q_{\alpha,j})}$. 
The introduced dynamics preserves the averaged number of atoms in each band, i.e. $N_{\rm X,M}=\int |{\cal A}_{\rm X, M}|^2 d{\bf r}$ are constants. 

If however, the matching conditions are met, then the other types of the resonant four-wave interactions, 
resulting in change of the average number of atoms in the bands, can occur.  Now the Hamiltonian governing 
evolution of the modulation of the Bloch states reads
\begin{eqnarray}
\label{ham2}
	H=\frac 12 \int d{\bf r} \left\{\left|\hat{M}_{\rm X}^{-1}\nabla {\cal A}_{\rm X}\right|^2+
\left|\hat{M}_{\rm M}^{-1}\nabla {\cal A}_{\rm M}\right|^2+\chi_{_{\rm X}}\left|{\cal A}_{\rm X}\right|^4
\right.
\nonumber 
\\
 \left.
 +\chi_{_{\rm M}}\left|{\cal A}_{\rm M}\right|^4+
 4\chi
 \left|{\cal A}_{\rm X}\right|^2 \left|{\cal A}_{\rm M}\right|^2+\chi{\cal A}_M^2\bar{\cal A}_X^2+
\chi\bar{\cal A}_M^2{\cal A}_X^2
 \right\}.
\end{eqnarray}

The resonant processes occur simultaneously with the nonresonant transitions between states in different 
bands but with the same wavevector (as those observed in~\cite{kivshar}). The latter, however, are negligible 
in our configuration. This was proved numerically by evaluating projections of the obtained distributions into the states $\Psi_{\rm M, X}$.

It follows from (\ref{ham1}) and  (\ref{ham2}) that stability of initial distributions is determined by the signs of $M^{-1}_{jj}$~\cite{BKS}. Since we consider the 
condensate with a positive scattering length,  and $M^{-1}_{jj}({\bf q}_{\rm X})>0$ and 
$M^{-1}_{jj}({\bf q}_{\rm M})<0$ ($j=x,y$),  
in the both cases the states $\Psi_{\rm M}$ and $\Psi_{\rm X}$ are unstable and stable, respectively,   since one of the stability conditions of a band $\alpha$ populated with the density $\rho_\alpha$ reads $Z(Z+2\chi_\alpha\rho_\alpha^2)<0$ with $Z=q_x^2/M_x+q_y^2/M_y$. Meantime, in the resonant case  (\ref{ham2}) there exists an additional domain of instability  defined by the condition $[Z+\rho_\alpha^2(3-\chi_\alpha)][Z+\rho_\alpha^2(1-\chi_\alpha)]<0$, what results in faster development of instability, reported in numerical simulations below.
Thus, tunneling can either stabilize or destabilize the dynamics of the wave function  depending on the direction of the tunneling that corresponds to M$\to$X and X$\to$M processes, respectively.  

\begin{figure}[h]
\epsfig{file=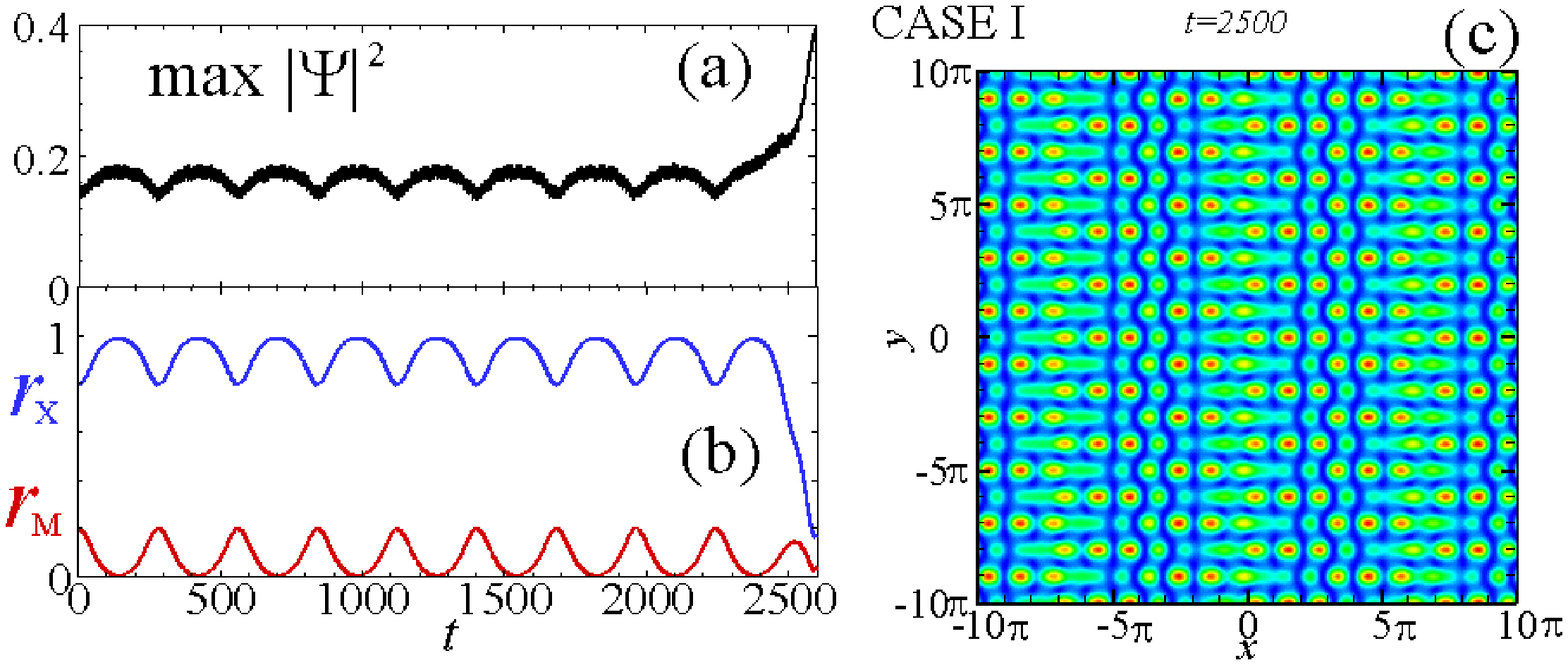,height=8cm,angle=-90}
\epsfig{file=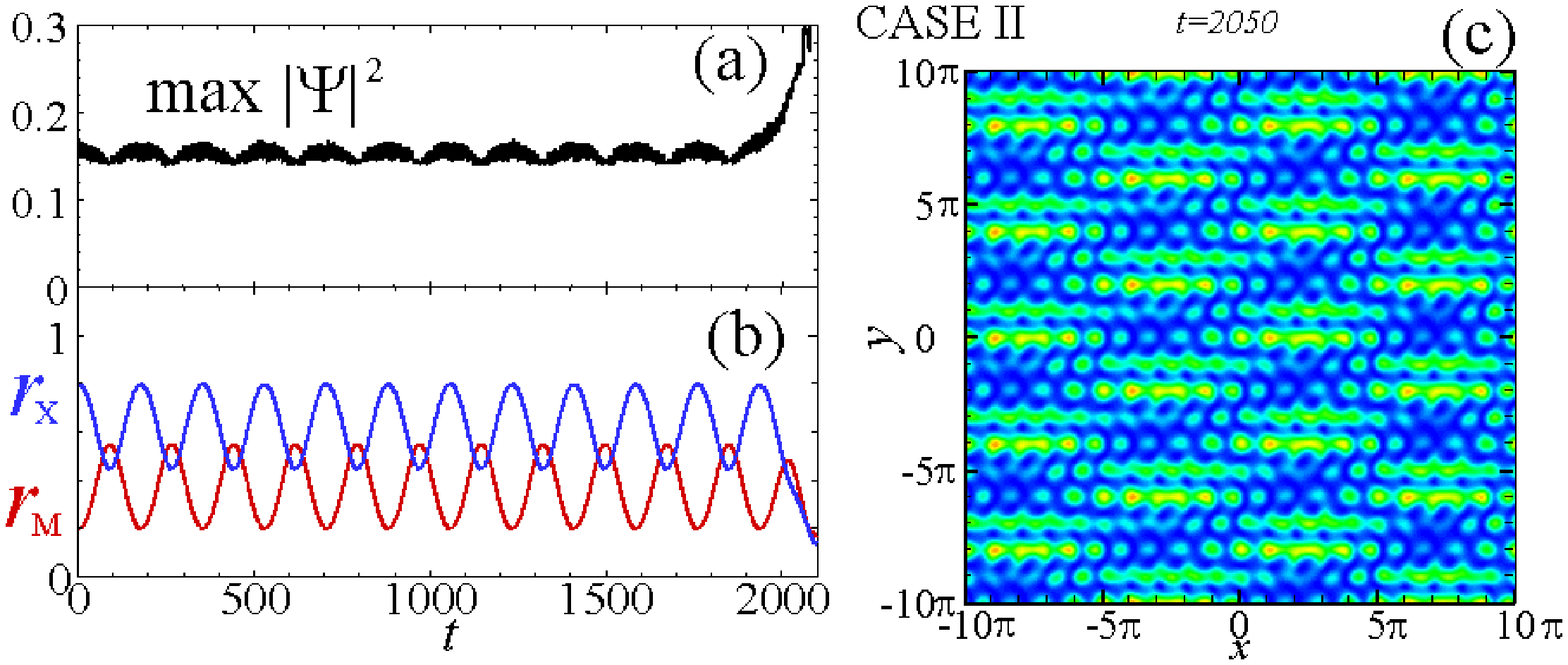,height=8cm,angle=-90}
\epsfig{file=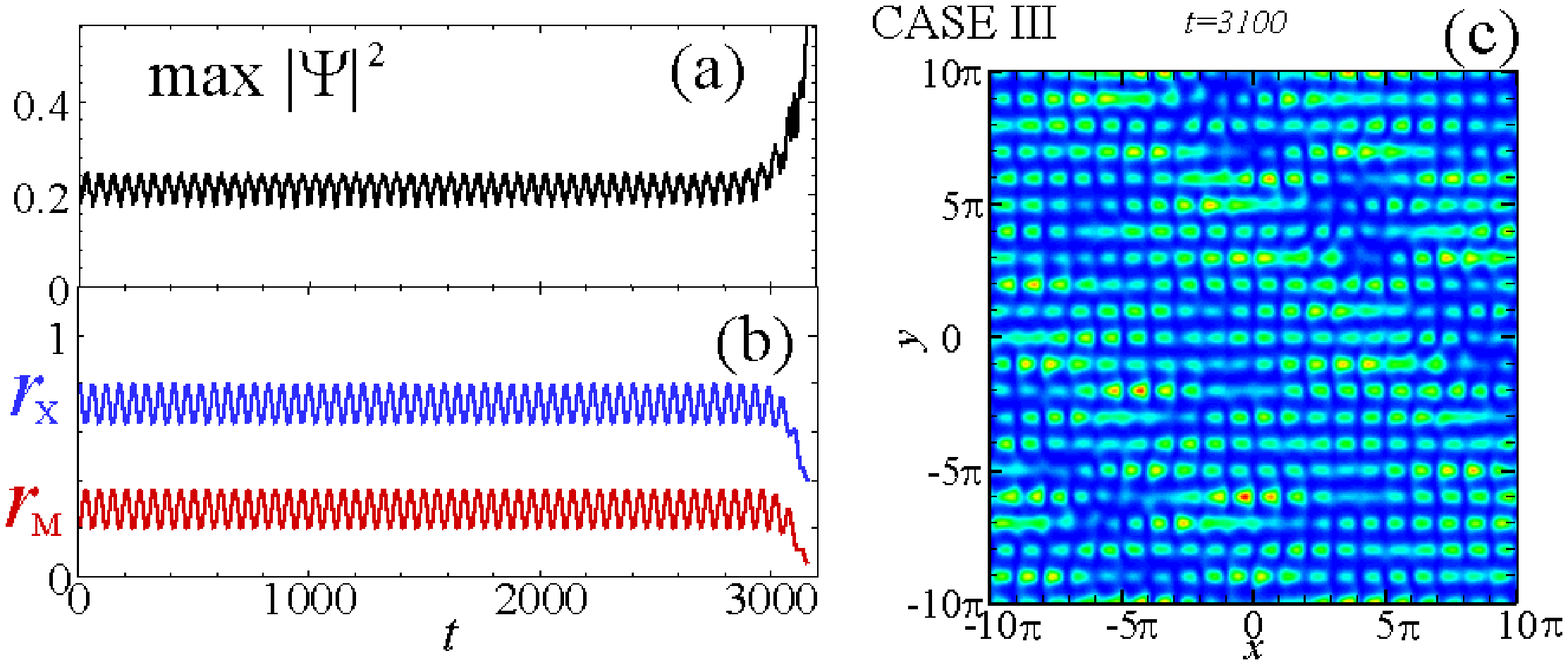,height=8cm,angle=-90}
\caption{[Color online] (a) Dynamics of the density maximum; (b) Time evolution of the occupation rates $r_{\rm M}(t)$ 
(red lines) and $r_{\rm X}(t)$ (blue lines). (c) Snapshots of the atomic density distributions after instability 
is developed. All simulations are done with the initial distributions $r_{\rm M}(0)=0.2$ and $r_{\rm X}(0)=0.8$.
The cases I, II, and III correspond to those in Fig.~\ref{fig1}.
}
\label{fig2}
\end{figure}

\begin{figure}[h]
\epsfig{file=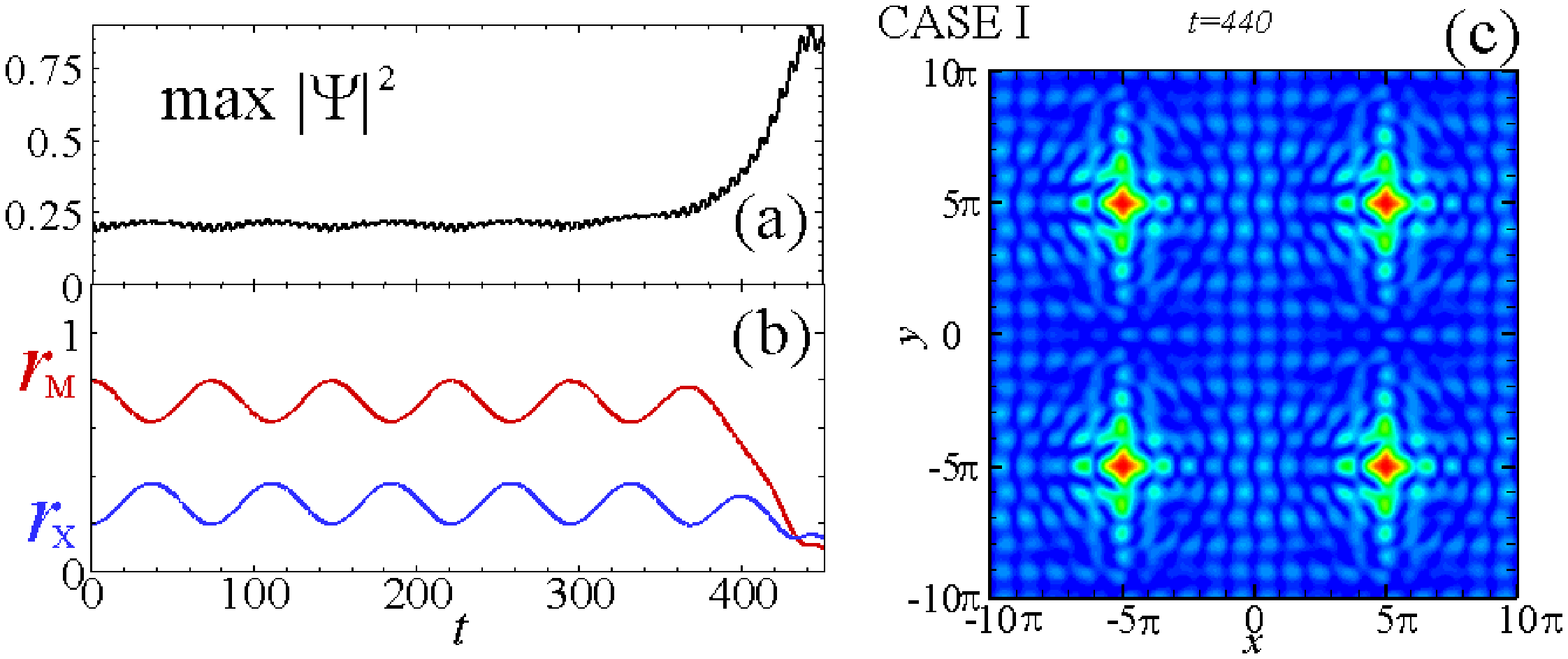,height=8cm,angle=-90}
\epsfig{file=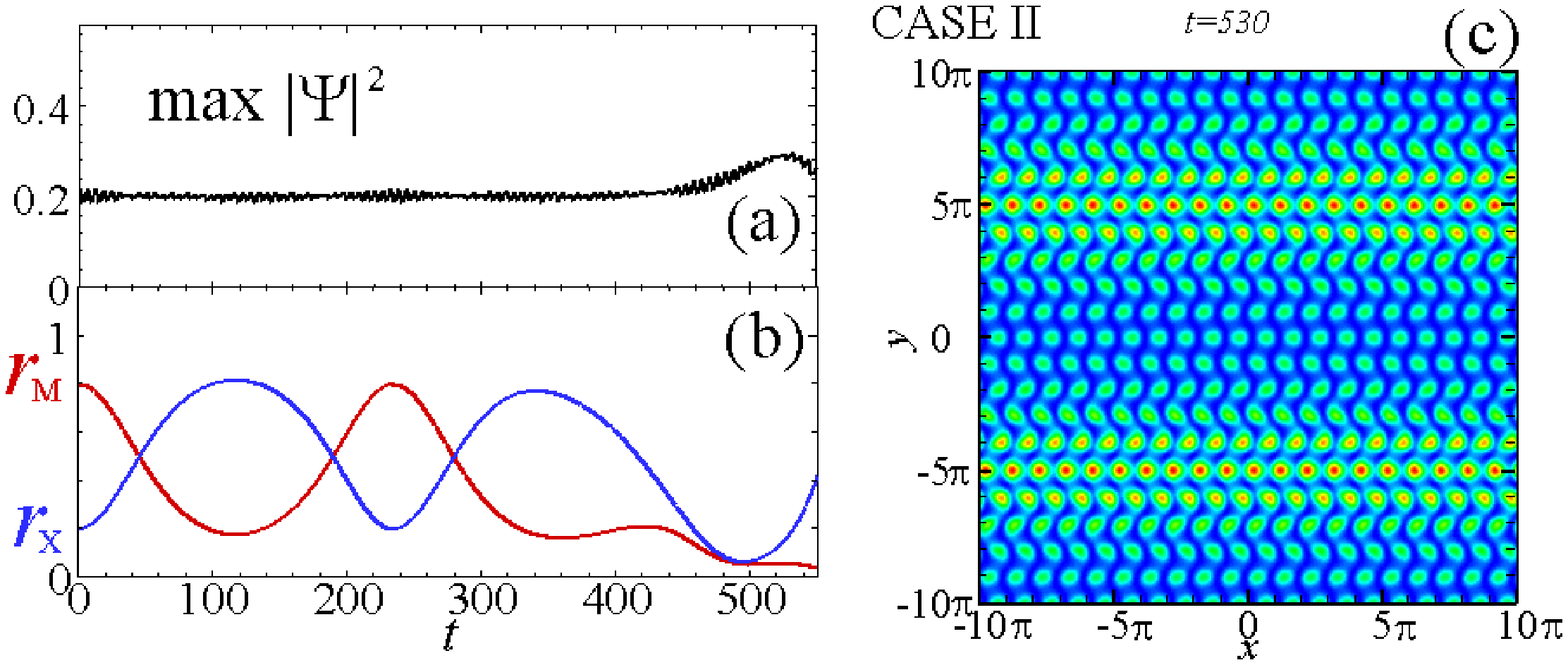,height=8cm,angle=-90}
\epsfig{file=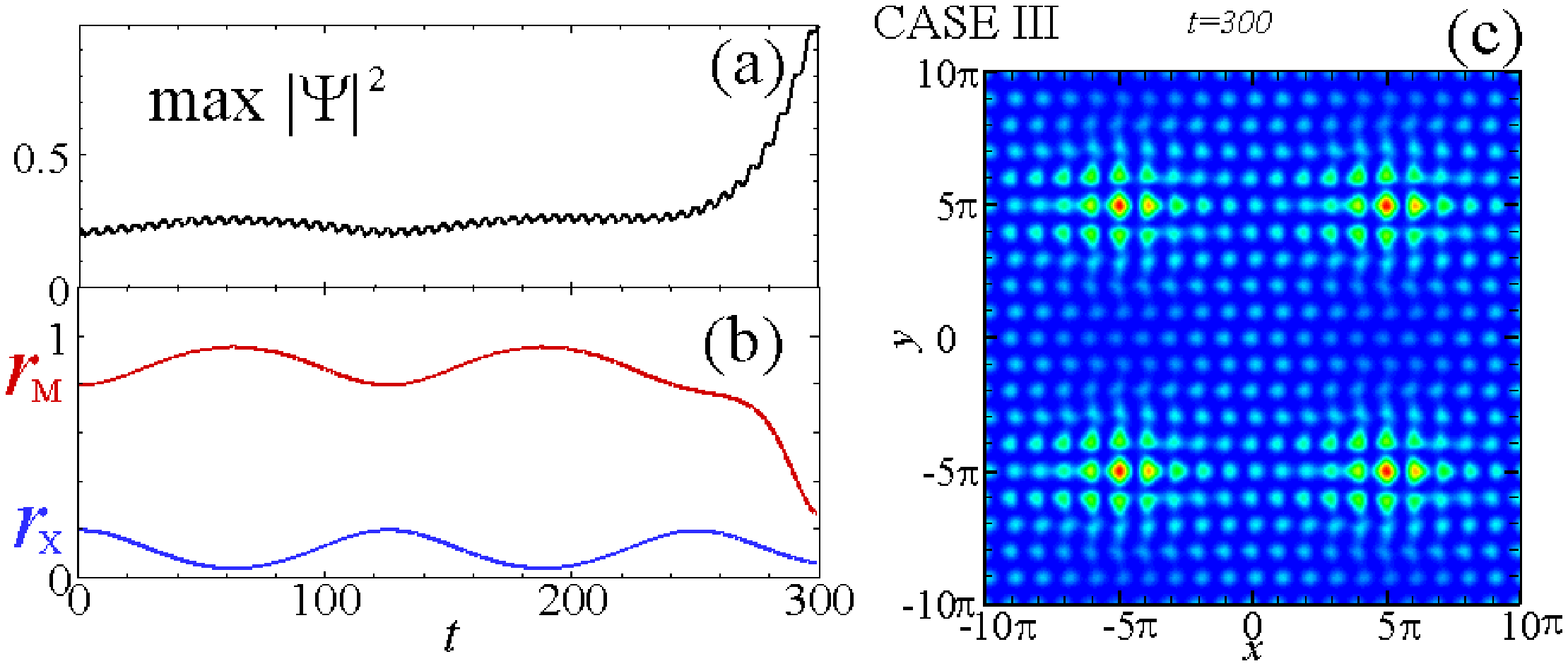,height=8cm,angle=-90}
\caption{[Color online] The same as in Fig.\ref{fig2}, but with initial particles distribution between points M and X taken as follows: $r_{\rm M}(0)=0.8$ and $r_{\rm X}(0)=0.2$. To accelerate modulational instability  we impose an initial 
perturbation $0.1 |\Psi_{\max}|\sin\left(0.314 x\right)\sin\left(0.314 y\right)$ on the numerically obtained 
linear Bloch states.}
\label{fig3}
\end{figure}

To inspect the interplay between tunneling and instability, we carried out direct numerical simulations 
of Eq.~(\ref{GPE}), subject to periodic boundary conditions and initial condition in a form 
$\alpha_{\rm M}\Psi_{\rm M}+\alpha_{\rm X}\Psi_{\rm X}$ [c.f. (\ref{psi})] where the factors $\alpha$ 
determine the distribution of atoms between the states. Simultaneous use  of smooth initial perturbation 
allowed us to accelerate the development of the instability (c.f. time scales in Fig.~\ref{fig2}, where no 
initial perturbation is applied and Fig.~\ref{fig3} where initial states were perturbed).
For quantitative analysis we computed the occupation rates $r_{\rm X,M}=N_{\rm X,M}/N$, where 
$N=N_{\rm X}+N_{\rm M}$ is the total number of atoms.
When the two states have nonzero initial population, the nonlinearity results in periodic particle exchange 
between these states. When tunneling is nonresonant (Cases I and III in Figs.~\ref{fig2},~\ref{fig3}) this 
exchange is relatively weak: it is caused by wave interactions of higher orders not accounted by the 
Hamiltonian (\ref{ham1}). Exchange of atoms between the bands becomes the leading order phenomenon when 
tunneling is resonant (Case II) what is described by additional hopping integrals in (\ref{ham2}). 
 
Let us consider X$\to$M process shown in Fig.\ref{fig2}. Period of oscillations $T$ decreases as amplitude potential 
$|V_0|$ increases: from the top to bottom $T\approx 560,\,362,\,116$. This is explained by growth of the cross-phase 
modulation: (from top to bottom) $\chi \approx 6.4;6.8;7.6 \times 10^{-3}$. When the gap is opened (Case~III) the 
system stays for a long time in a stable position due to low occupation of the unstable state M. Nevertheless, because 
of periodic increase of the population of the state $M$  the instability is developed at $t\approx 3100$. Possibility 
of tunneling accelerates this process: in the  Case~I the instability is developed at $t\approx 2500$. The most 
substantial enhancement of the instability is observed when the tunneling becomes resonant: in the Case II the 
instability occurs at $t\approx 2050$.  The development of the instability results in emergence of coherent 
structures. If tunneling is resonant (Case II) atomic distribution displays asymmetric peaks, which evolve in 
time. The peaks are elongated in $x$-direction, what seems to contradict to an intuitive assumption that the 
symmetry of the developed coherent structure resembles the symmetry of the unstable state, from which they are 
evolved~\cite{BKS}. Such an interpretation well corroborates with the results shown below in the Cases~I and III 
of Fig.\ref{fig3}. In the case at hand, however, one has to keep in mind that, while the tensor of the inverse 
effective mass is symmetric at the unstable point M, it is not symmetric any more at the point X.  Specifically, 
for the Case~II in Fig.~\ref{fig2}  $\hat{M}_{\rm X}^{-1}\approx\,$diag$(.3,.03)$. Since $M^{-1}_{xx}>M^{-1}_{yy}$, 
when the atoms are at the X-point, dispersion of the wave along the $x$-direction is stronger than the dispersion 
along $y$-axis. Since instability is developed due to particle exchange between the two states, this explains 
elongation of the emerging excitations along the $x$-axis. 

Now we turn to the situation when initially the most of particles occupy unstable state M, i.e. to the M$\to$X 
process, illustrated  in Fig.~\ref{fig3}. In the configuration where there is no linear tunneling (Case~III) 
the instability is developed at $t\approx 270$. Nonresonant tunneling (Case~I)  leads to some attenuation of 
the instability that starts at $t\approx 370$. In the both cases the developed coherent structures resemble 
those observed in Ref.~\cite{BKS} for a separable square lattice and reflect the symmetry of the lattice. 
Dramatical changes occur in the case of the resonant Bloch-band tunneling (the Case~II). Besides significant 
attenuation of instability, which now occur at $t\approx 450$, one observes change of the symmetry of the 
developed structure. Like in the case of X$\to$M tunneling, the emerging pulses are strongly localized along 
$y$-axis and elongated along $x$-axis, what corroborates with the fact that the attenuation of the instability 
has occurred due to tunneling into the stable state X.
 
Different symmetries of the emerging structures, that can be identified e.g. by means of diffraction imaging, 
provide a possibility for the experimental observation of the phenomenon and identification of the type of 
the process. We also mention that although the time difference in development of instability differs in our 
numerics only by 10\%$\div$30\% for different processes, this is experimentally observable time. Indeed, for 
condensed rubidium atoms in calculations 50 units of the dimensionless time correspond to 2 ms, what means 
that time difference in instability of different regimes is of order of 8$\div$20 ms.

The results shown in Figs.~\ref{fig2},~\ref{fig3} reveal a common feature, namely that until the  instability  
is developed the atoms are mainly distributed between the two interconnected states $\Psi_{\rm M}$ and 
$\Psi_{\rm X}$. This is a strong support of the validity of a two-mode approximation suggested in \cite{KKS}, 
and in this Letter defined by the Hamiltonians (\ref{ham1}) and (\ref{ham2}). However, after the instability 
is developed, we observed transfer of atoms to states other than $\Psi_{\rm X,M}$ (see panels (b) in Figs~\ref{fig2},~\ref{fig3}).  When such transfer takes place the two-mode approach is not applicable any more.

To conclude, we have shown that two-body interactions in a BEC loaded in a two-dimensional orthogonal nonseparable optical lattice result in a strong interplay between the phenomena of Bloch-band tunneling and modulational instability. As a consequence, tunneling appears to be one of the key factors  for the superfluid-insulator transition, especially in the case of lattices with negligible gap, where matching conditions for the four wave interactions are provided. The gap width, meantime, is 
an easily changeable lattice parameters and can be controlled by the laser intensity. Specifically we 
found that the atoms, initially loaded in a stable state, due to inter-band exchange   develop instability, which is significantly enhanced when the matching 
conditions are satisfied. On the other hand resonant tunneling attenuates the instability when atoms are initially loaded in an unstable state. We also found that the instability makes the Bloch-band tunneling strongly asymmetric in the sense of dependence of developed patterns on initial atomic distribution.
 
VVK acknowledges Yu. S. Kivshar for drawing our attention to 
Ref.~\cite{kivshar}. V.A.B. was supported by the FCT grant SFRH/BPD/5632/2001. VAB and VVK were 
supported by the FCT and FEDER under the grant POCI/FIS/56237/2004. Collaboration was supported by the 
bilateral Agreement GRICES/Czech Academy of Sciences and by COST P11 Action.

\end{document}